\documentclass[prb,aps,twocolumn,superscriptaddress,showpacs]{revtex4}

\usepackage{graphicx}

\begin{document}

\title{High energy neutron scattering from hydrogen using a direct geometry spectrometer}

\author{C. Stock}
\affiliation{ISIS Facility, Rutherford Appleton Labs, Chilton, Didcot, OX11 0QX, UK}

\author{R.A. Cowley}
\affiliation{Oxford Physics, Clarendon Laboratory, Parks Road, Oxford, United Kingdom OX1 3PU, UK}
\affiliation{Diamond, Rutherford Appleton Laboratory, Chilton, Didcot, OX11 0QX, UK}

\author{J.W. Taylor}
\affiliation{ISIS Facility, Rutherford Appleton Labs, Chilton, Didcot, OX11 0QX, UK}

\author{S.M. Bennington}
\affiliation{ISIS Facility, Rutherford Appleton Labs, Chilton, Didcot, OX11 0QX, UK}

\date{\today}

\begin{abstract}

Deep inelastic neutron scattering experiments using indirect time-of-flight spectrometers have reported a smaller cross section for the hydrogen atom than expected from conventional scattering theory.  Typically, at large momentum transfers, a deficit of 20-40\% in the neutron scattering intensity has been measured and several theories have been developed to explain these results.  We present a different approach to this problem by investigating the hydrogen cross section in polyethylene using the direct geometry time-of-flight spectrometer MARI with the incident energy fixed at a series of values ranging from E$_{i}$=0.5 eV to 100 eV.  These measurements span a much broader range in momentum than previous studies and with varying energy resolutions.  We observe no momentum dependence to the cross section with an error of 4\% and through a comparison with the scattering from metal foil standards measure the absolute bound cross section of the hydrogen atom to be $\sigma(H)$= 80 $\pm$ 4 barns.  These results are in agreement with conventional scattering theory but contrast with theories invoking quantum entanglement and neutron experiments supporting them.  Our results also illustrate a unique use of direct geometry chopper instruments at high incident energies and demonstrate their capability for conducting high-energy spectroscopy.

\end{abstract}


\maketitle

\section{Introduction}

\begin{figure}[t]
\includegraphics[width=75mm]{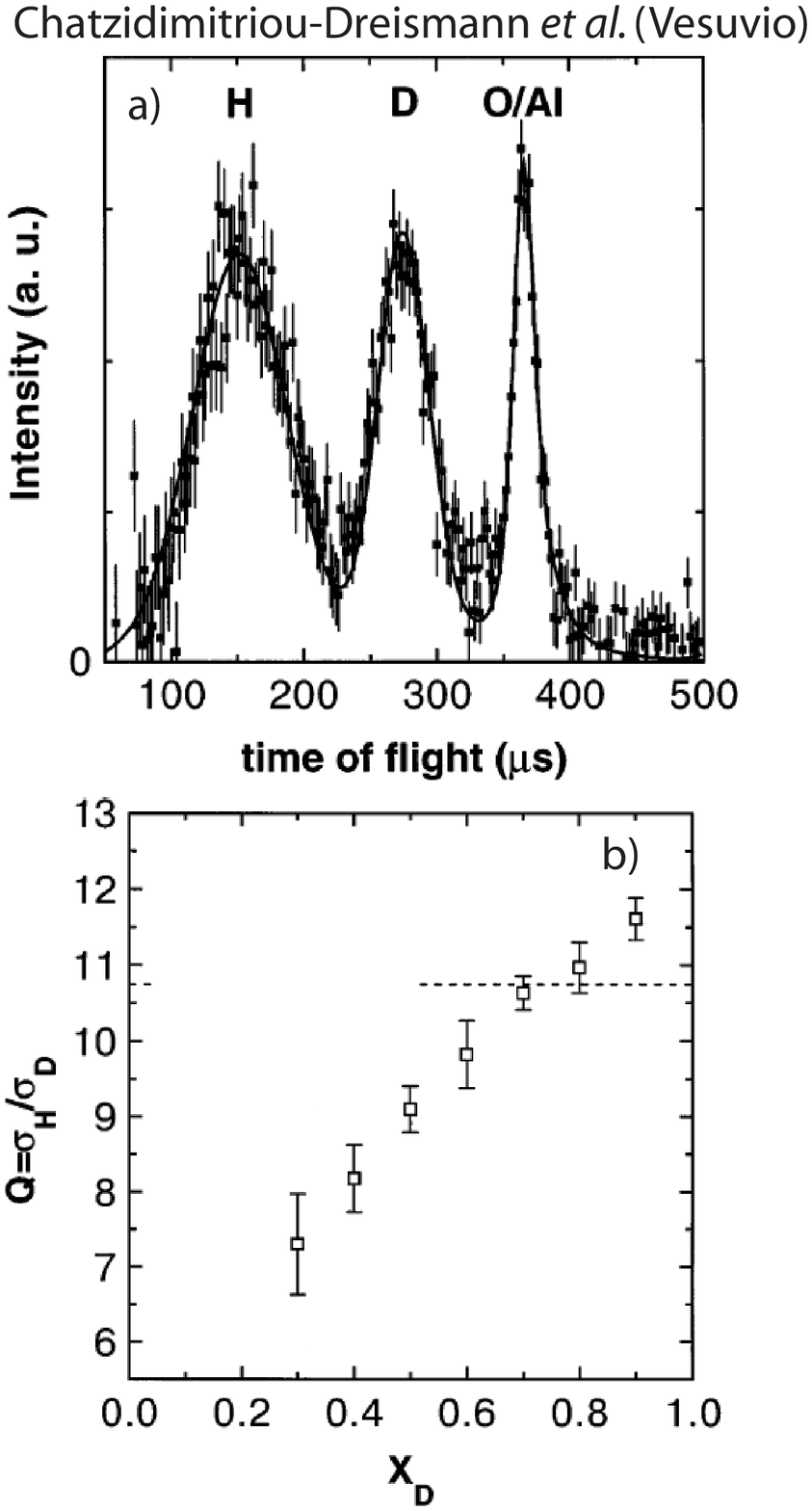}
\caption{The initial results illustrating an anomalous deficit of intensity in H$_{2}$O and D$_{2}$O.  The time-of-flight signal is plotted in panel $a)$ which illustrates the distinct separation observed between the $H$ and $D$ recoil peaks.  The ratio of the intensities as a function of D$_{2}$O concentration is shown in panel $b$.  The dashed line at 10.7 is the expected value based on conventional scattering theory.  The results are taken from Ref. \onlinecite{Dreis97:79}.} \label{water}
\end{figure}

There has been great interest in the neutron scattering cross section from hydrogen since the experiments of Chatzidimitriou-Dreismann \textit{et al}.~\cite{Dreis97:79}  This initial work investigated the scattering from hydrogen in mixtures of H$_{2}$O and D$_{2}$O and found an anomalously low value of the hydrogen cross section.  The results are summarized in Fig. \ref{water} where the ratio of the hydrogen and deuterium cross sections ($\sigma(H)/\sigma(D)$) as a function of deuterium content is illustrated in panel $b)$ and panel $a)$ illustrates a representative time-of-flight spectra from which the cross sections were derived.  The dashed line in Fig. \ref{water} $b)$ represents the expected value of $\sigma(H)/\sigma(D)$=10.7 based on conventional scattering theory.  The data clearly deviate from expectations and were initially explained in terms of entanglement between hydrogen and deuterium atoms on very short (attosecond) timescales.  

Since this initial experiment, similar hydrogen cross section deficits have been observed in metal hydrides and hydrogen containing metallates (Refs. \onlinecite{Karlsson03:67,Krzystyniak07:126,Karlson05:779,Redah00:276,Reis02:330,
Karlsson99:46,Karlsson02:74,Redah06:385}), polystyrene and benzene (Ref. \onlinecite{Dreis02:116}), Formvar (Ref. \onlinecite{Dreis03:91}) and HCl (Ref. \onlinecite{Senesi05:72}).  All of these results have been obtained with high-energy neutron scattering at large momentum transfers using indirect geometry instruments (in particular the Vesuvio/eVs spectrometer at ISIS). The results, if correct, would represent a significant failure of conventional models of scattering and several theories have been developed to explain these results including the breakdown of the Born-Oppenheimer approximation (Ref.\onlinecite{Gido05:71,Reiter05:71,Lin04:69}) and quantum entanglement (Ref. \onlinecite{Karlsson99:46,Karlsson00:61,Karlsson02:65,Karlsson03:90}) at short timescales.  The latter theory has been refuted in Refs. \onlinecite{Cowley03:15,Sugimoto05:94,Sugimoto06:73}, where it has been shown to be inconsistent with the first-moment sum rule and also argued that good energy resolution and low temperatures are required to observe the effects suggested to  result from entanglement.  Such a sum rule violation does not occur in the former theories where a redistribution of intensity conserves the first-moment while resulting in an apparent decrease in intensity of the hydrogen scattering.  However, Ref. \onlinecite{Colognesi05:358} claims the former theories give too small an effect to be consistent with experiments.

While supporting evidence has been claimed from electron scattering (Ref. \onlinecite{Cooper08:100}), Raman spectroscopy (Ref. \onlinecite{Dreis95:75}), and neutron diffraction (Ref.\onlinecite{Fillaux06:18}), several noteworthy attempts using neutron scattering have failed to reproduce a deficit in the cross section.  Neutron interferometry has been conducted on mixtures of H$_{2}$O and D$_{2}$O and have found no deviations from expectations based on two fluids.~\cite{Ioffe99:82}  Several indirect geometry neutron scattering measurements utilizing the foil filter technique, an experimental setup similar to the initial study presented in Ref. \onlinecite{Dreis97:79}, have been conducted and report no evidence for a missing hydrogen cross section.~\cite{Moreh05:94,Moreh06:96,Blostein09:102}  These studies have been reconciled with the initial data based on the large difference in the energy resolution and hence have questioned whether ${d \sigma/d\Omega}$ or ${d^{2}/dE d\Omega}$ was being measured and compared.~\cite{Karlsson04:92,Mayers07:98,Dreis00:84}  A detailed revisiting of the initial experimental setup on polyethyelene did find evidence for an anomalous lack of intensity consistent with previous results, however several experimental concerns were also noted.~\cite{Cowley06:18}

Given the large discrepancy in experimental results and the theoretical interest in this effect, we have pursued a different approach to the investigation of the hydrogen cross section as a function of momentum transfer.  While all previous experimental studies have been conducted with indirect geometry spectrometers (where the final neutron energy is fixed at one energy), we present a direct geometry study of the cross section of hydrogen in polyethylene using the MARI chopper spectrometer.  Because the incident beam energy can be fixed at a particular value, the hydrogen cross section can be investigated over a range of momentum transfer with varying energy resolution and this allows a direct comparison with the experiments and theories described above.  This experiment also allows overlapping data to be obtained in both momentum and energy transfer thereby checking the consistency of the experimental results.  This feature of direct geometry spectrometers is not possible on indirect machines where a filter edge or analyzer forces one particular fixed final energy.  

Given the previous results on all hydrogen containing materials, there are two key results which need to be investigated and provide the basis for the experiment reported here.  Firstly, the momentum dependence of the scattering line shape needs to be measured over a broad range extending to over $Q\sim 100\AA^{-1}$.  Secondly, the absolute value (in units of barns) is required for these momentum transfers and comparison with the theoretical value made.  

We will show that within experimental error, there is no change in the hydrogen cross section with momentum transfer and that the absolute value of the cross section is consistent with conventional scattering theory.  These results show that conventional scattering theory based on the Born-Oppenheimer approximation and the impulse approximation is valid and provides a good description of the data.  These results also demonstrate the use of direct chopper instruments for conducting spectroscopy at high energies and we suggest possible future uses of these instruments for understanding problems in magnetic and strongly correlated electronic systems.

The paper is divided into five sections and two appendices.  The first section discusses the experimental setup including the direct geometry MARI spectrometer and also how we have measured and subtracted the background.  The second section reviews the impulse approximation and the data analysis which provides the basis for understanding the experimental results.  We also review the kinematics of a direct geometry spectrometer.  The third section describes the calibration of the spectrometer using a series of ``standard'' metal foils such that the experiments could be put on an absolute scale.  The fourth section describes the hydrogen cross section and outlines the momentum dependence and the absolute value obtained from the comparison with the results from the metal foils.  We end the paper with a section of discussion, conclusions and future plans.   The appendices include discussions on the energy calibration and resolution.

\section{Experiment}

\subsection{The MARI direct geometry spectrometer}

\begin{figure}[t]
\includegraphics[width=90mm]{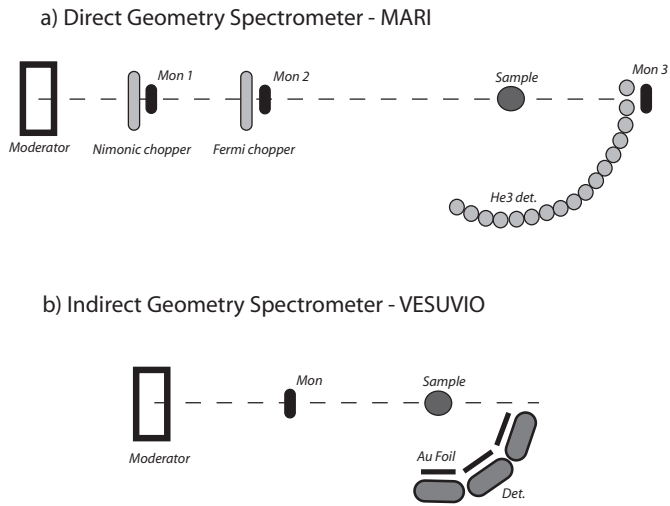}
\caption{A schematic of the direct geometry ($a)$ MARI and indirect geometry ($b)$ VESUVIO spectrometer.  $a)$ The locations of the beam monitors (labeled Mon 1 etc.), nimonic and Fermi choppers, and detectors are marked.  The incident beam energy is fixed by the Fermi chopper and scattering neutrons are detected by the He$^{3}$ detectors.  $b)$ No Fermi chopper exists on VESUVIO and an incident beam with a spectrum of energies is incident on the sample.  The final energy is fixed by using an Au foil resonance.} \label{spectrometers}
\end{figure}

MARI is a direct geometry time-of-flight spectrometer as shown in Fig. \ref{spectrometers} $a)$. The spectrometer consists of a Fermi chopper 10.02 m from the neutron moderator, a sample position 1.77 m from the chopper and a detector bank of He$^{3}$ tubes located 4.02 m underneath the sample. The incident energy is determined by the phasing of the Fermi chopper with respect to the neutron pulse produced in the moderator while the scattered energy is determined by the time-of-flight of the scattered neutrons.

For energy transfers below $\sim$ 1 eV, both the nimonic and Fermi chopper are utlized. The nimonic chopper removes high-energy neutrons (typically above $\sim$ 3-5 eV) and other high-energy radiation that is emitted from the moderator at very early times after the arrival of the proton pulse on the target. For our high energy neutron experiments with incident energies above 1 eV, the nimonic chopper was removed so that the high-energy spectrum could be accessed. This inevitably resulted in a significant background at short times on the detectors.  The subtraction of this background and spectrometer calibration are discussed later.

The Fermi chopper consisted of blades of neutron absorbing boron nitride.  The blades were curved such that for incident energies of 500 meV, the neutrons would have an unobstructed path when the chopper is spinning at 600 Hz.  Various other blade configurations are available for lower incident energies.  The incident energy was fixed by adjusting the phase between the opening of the Fermi chopper and the time the proton pulse reached the target.  The incident energy was further calibrated for the effects of drift in the detectors and possible timing errors by recording the time the neutron pulse reached the $Mon$ 2 and $Mon$ 3 glass bead detectors.  This calibration is outlined in detail in Appendix A to this paper.  Because the distances are all known, a time-of-flight signal in both these monitors provides an unambiguous measurement of the incident neutron energy.  $Mon$ 1 was used to normalise all data sets to an equal incident beam flux by dividing by the integrated counts over all times-of-flight. 

The detectors were cylindrical tubes of He$^{3}$ at 10 atm of pressure.  The detector efficiency as a function of energy was calculated based on the known cross-section of He$^{3}$ assuming a cylindrical geometry.~\cite{Erik_private}  Counts were recorded on the detector with a constant time intervals of $\delta t$=0.25 $\mu s$ up to 1000 $\mu s$ from when the proton pulse is incident on the target.  To convert counts measured on the detectors to absolute units (barns) we used a series of metal foils with known tabulated cross sections.  We describe the method of obtaining the calibration constant later in this paper.

\begin{figure}[t]
\includegraphics[width=90mm]{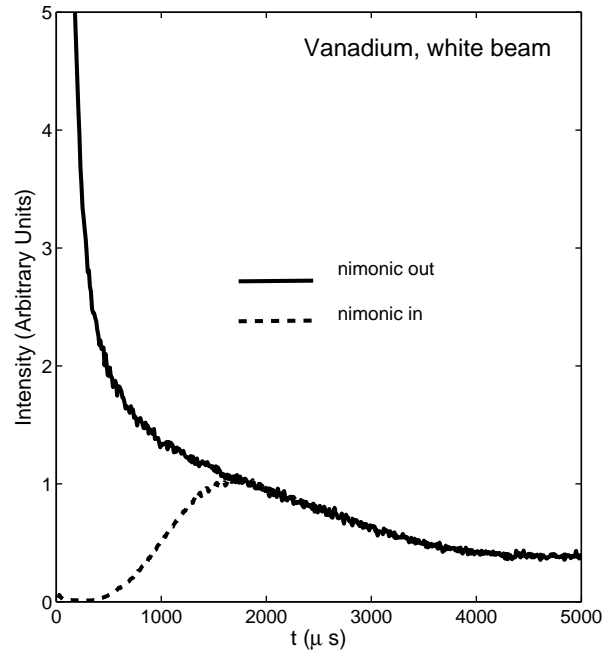}
\caption{The scattered intensity off a vanadium foil with a white beam of neutrons incident on the sample as measured on the detectors.  The dashed line is with the nimonic chopper in and the solid line is with the nimonic removed.  The data is integrated for $2\theta=[5^{\circ},130^{\circ}]$.} \label{nim_wb}
\end{figure}

We conducted two different experiments. The first measured the scattering of polyethylene wrapped around an aluminum annulus.  Given the need to obtain absolute cross sections for the hydrogen atom and difficulties in removing the aluminum intensities, we performed a second experiment using samples of either low-density (LDPE) or high-density (HDPE) polyethylene (CH$_{2}$) films.  The films and metal foils were free standing and mounted  at 42$^{\circ}$ to the vertical axis.  This setup had the advantage that only the sample was in the beam for any given measurement and this simplified the analysis considerably without the need to consider the presence of aluminum in the scattered signal.   Both experiments gave consistent results and we focus on the second experiment based on free standing films in this paper. The physical properties (including thickness $d$, density $\rho$, and molar mass $m$) of the polyethylene samples are listed in Table \ref{table_poly}.

\begin{table}[ht]
\caption{Polyethylene Samples}
\centering
\begin{tabular} {c c c c}
\hline\hline
X & d (mm) & $\rho$ (g/cm$^{3}$) & $m$ (g/mol) \\
\hline\hline
HDPE& 0.044 & 0.95 &  14.03 \\
LDPE1 & 0.20 & 0.92 &  14.03 \\
LDPE2 & 0.40 & 0.92 & 14.03 \\
\hline
\label{table_poly}
\end{tabular}
\end{table}

It is interesting to compare the MARI direct geometry spectrometer described here with VESUVIO which was used for the earlier experiments on hydrogen. VESUVIO is an inverse time-of-flight machine which has a pulsed white incident beam of neutrons that are scattered from the sample and the scattered energy is determined by the transmission through a thin film for which there is a nuclear resonance such as gold. The energy of the incident neutrons is known from the time-of-flight of the neutrons and energy of the scattered neutron is determined by the nuclear resonance. This spectrometer has the disadvantage that it involves the subtraction of counts with the gold foil `in' and the foil `out'. It also depends crucially on the energy dependence of the incident neutron flux. Since there are very few metal foils that have suitable resonance energies and widths for use in this type of spectrometer (outlined in Ref. \onlinecite{Imberti05:552}) it is not possible to vary appreciably the scattered neutron energy and so test the instrumental results by measuring a particular wavevector and energy transfer with different neutron energies. In contrast, a direct time-of-flight instrument such as MARI has a monochromatic incident beam and the detectors measure neutrons with a range of different energies. The resultant flux is then crucially dependent on the knowledge of the efficiency of the detectors as a function of the neutron energy. The incident energy for this type of instrument can, however, be controlled and varied and this enables experiments to be performed to ensure that measurements at different incident energies give consistent results.

\subsection{Background characterisation and subtraction}

As described in the previous section, because of the requirement of large incident energies the nimonic chopper was removed for these studies.  This had the effect that (while allowing access to high-energy neutrons) it also allowed high-energy charged particles and $\gamma$-rays from the initial interaction of the proton pulse with the tungsten target to be incident on the sample.  This contributed to a significant increase in the background especially at short times of flight.  This is shown in Fig. \ref{nim_wb} which illustrates the counts measured on the detectors integrated over $2\theta=[5^{\circ},130^{\circ}]$ with the nimonic chopper in and out.  The Fermi chopper was removed for these measurements to ensure a white beam spectrum incident on the sample position. The sample was a vanadium foil of thickness 0.075 mm.

\begin{figure}[t]
\includegraphics[width=80mm]{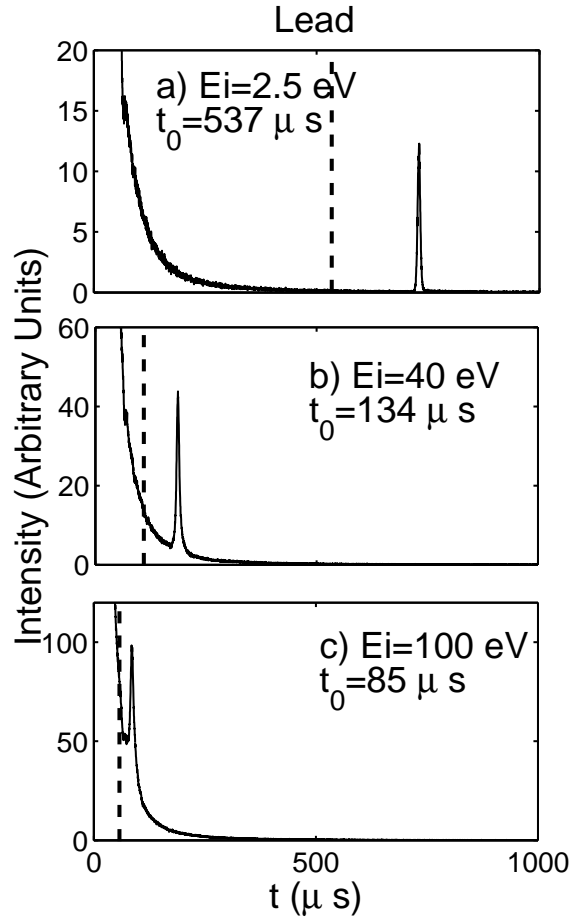}
\caption{Time of flight spectra for a lead foil sample for E$_{i}$=2.5 eV, 40 eV, and 100 eV.  The data is integrated over $2\theta$=[30$^{\circ}$,70$^{\circ}$].  $t_{0}$ represents the time the pulse of neutrons strikes the sample and is graphically represented by the vertical dahsed lines.} \label{back_fig}
\end{figure}

\begin{figure}[t]
\includegraphics[width=80mm]{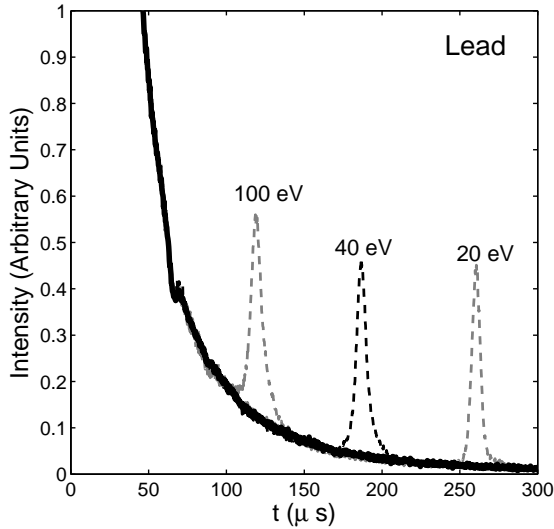}
\caption{The time-of-flight spectra for lead foil plotted for E$_{i}$=20, 40, and 100 eV.  The background derived from the method described in the text is plotted as the solid and thick black line.} \label{back_example}
\end{figure}

\begin{figure}[t]
\includegraphics[width=90mm]{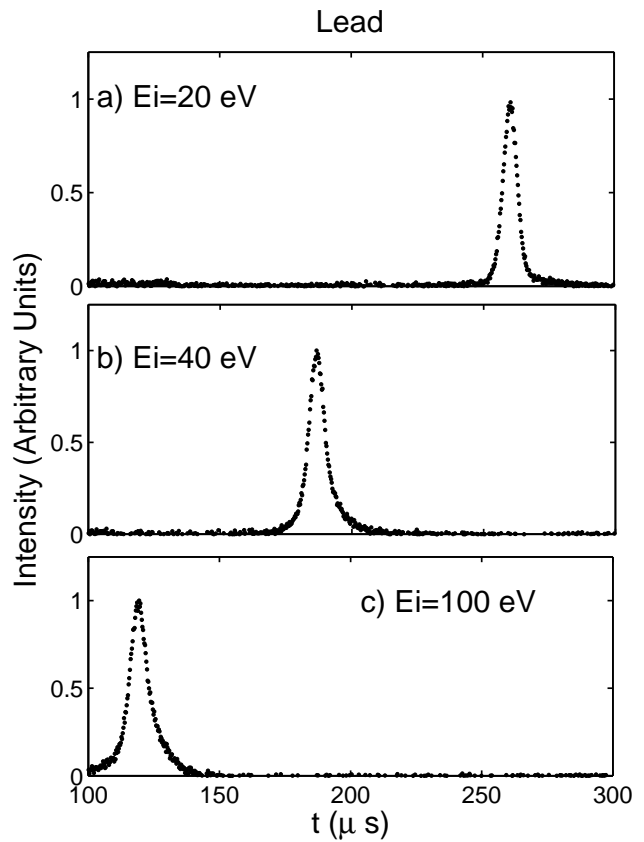}
\caption{The time-of-flight spectra with a lead foil placed at the sample position and with the background subtracted by the method described in the text.  The intensities were obtained by integrating $2\theta$=[30,70]. The background is zero within the errors at times both shorter and larger than the peaks in the spectra.} \label{lead_time}
\end{figure}

Fig. \ref{nim_wb} illustrates that at small times, below $\sim$ 1000 $\mu s$, the nimonic chopper has a significant effect on the spectrum and counts on the detector.  A particuarly strong feature is a significant tail which seems to diverge as $t$ approaches $t=0$.  It is essential to understand this strong background scattering so that it can be quantitatively subtracted to obtain the correct intensity and line shape of the scattering.  We found that the background is to a good approximation independent of the incident neutron energy chosen.  This is demonstrated in Fig. \ref{back_fig} which shows the scattering for several different incident energies from a lead foil.  The dotted vertical line shows the time when the neutron pulse reached the sample ($t_{0}$) and so for earlier times, there can be no effect of the incident energy.  Hence, then by performing scans at different energies the background spectra can be constructed as shown for lead foil in Fig. \ref{back_example}.  The result is then independent of the incident neutron energy although it depends on the sample so a similar procedure was followed for each sample studied.  Representative plots of the background subtracted time-of-flight spectra for lead are plotted in Fig. \ref{lead_time}.  The plots show that the background subtraction described above does accurately remove the background intensity as the data subtracts to zero for both shorter and longer times than the lead recoil peak.  We describe in the next section how this time-of-flight spectra was converted into an energy spectrum.

\section{Kinematics and the impulse approximation}

In this section we review the kinematic equations based on energy and momentum conservation and line shape predicted by the impulse approximation.  Having discussed how we obtained a background subtracted data set in time in the previous section, we will also outline the conversion to energy spectra which can be compared directly with theory.

\subsection{Kinematics and equations for detected neutron intensity}

Based on conservation of energy and momentum, the peak positions and the neutron cross sections can be calculated.  If we consider a beam of incident neutrons with a wavevector $\vec{k}_{0}$ which scatters off a sample of hydrogen atoms resulting in a scattered wavevector of $\vec{k}_{1}$, the energy transfered is related to the momentum of the recoiled hydrogen atom by $E=\hbar^{2}Q^{2}/2M_{p}$, where $M_{p}$ is the mass of the proton.  Making the approximation that the neutron has the same mass as the hydrogen atom, energy conservation results in $Q^{2}=k_{0}^{2}-k_{1}^{2}$.  Taking conservation of momentum gives $Q^{2}=k_{0}^{2}+k_{1}^{2}-2k_{0} k_{1} \cos(2\theta)$, with $2\theta$ defined as the angle between the incident and final neutron wavevector or the scattering angle.  Subtracting the energy and momentum conservation equations results in

\begin{eqnarray}
k_{1}=k_{0}\cos(2\theta),
\label{jac1}
\end{eqnarray} 

\noindent while adding gives $Q=k_{0}\sin(2\theta)$.  This latter equation illustrates that the maximum momentum and energy transfer occur when $2\theta$=90$^{\circ}$. 

The measured intensity of the scattered neutrons as a function of $2\theta$ needs to be corrected for several factors.  The scattered cross section per unit incident flux into solid angle $d\Omega$ and an energy interval $dE$ ($d^{2}\sigma/d\Omega dE$) for a monotonic system is proportional to $k_{1}/k_{0} S(Q,E)$ where $S(Q,E)$ is the van Hove scattering function.  The derivation of $S(Q,E)$ assumes an incident flux of neutrons falling on the sample. The equations and correction factors for an indirect geometry machine (such as VESUVIO) have been described in Refs.\onlinecite{Cowley03:15} and \onlinecite{Cowley06:18}.  Here we outline the appropriate equations and corrections for a direct geometry spectrometer.

The flux on the sample depends on the moderator and the properties of the Fermi chopper.  The flux from the moderator is proportional to $C(E_{0})dE_{0}$ where $C(E_{0})$ is approximately proportional to $1/E_{0}$.  While it is important to know the flux as a function of energy on a indirect geometry spectrometer where only the final energy is fixed, the details of the incident beam spectrum are not crucial to the use of a direct geometry spectrometer where the incident flux and energy are fixed through the phasing and frequency of the Fermi chopper.

Since the chopper spins at a constant speed the flux coming through the chopper is proportional to the opening time which is almost independent of the phasing and hence of the energy.  The chopper then lets neutrons through for a constant time interval and so the flux through the chopper is $C(E_{0})dE_{0}/dt \Delta t$ where $\Delta t$ is the chopper opening time.  Since $E_{0}=1/2 m v^{2}=1/2 m (\ell/t)^2$, $dE_{0}/dt=m \ell^{2}/t^{3}$ and hence is proportional to $E_{0}^{3/2}$.  These equations combined give the following for the cross section,

\begin{eqnarray}
{{d^{2}\sigma}\over{d\Omega dE}}=A C(E_{0}) E_{0}E_{1}^{1/2} S(Q,E) \Delta t,
\label{cs1}
\end{eqnarray} 

\noindent where $A$ is a constant of proportionality.

The detectors in a direct geometry spectrometer record intensity as a function of time and hence measure $d^{2}\sigma/d\Omega dt$ rather than $d^{2}\sigma/d\Omega dE$.  The conversion factor between the two is $dE_{1}/dt$ and has been calculated above to be proportional $E_{1}^{3/2}$.  These relations can be put together to give the following expression for the measured cross section

\begin{eqnarray}
{{d^{2}\sigma}\over{d\Omega dE}}={{d^{2}\sigma}\over{d\Omega dt}} \left({dE \over dt}\right)^{-1}={{d^{2}\sigma}\over{d\Omega dt}} {1 \over E_{1}^{3/2}},
\label{cs2}
\end{eqnarray} 

\begin{figure}[t]
\includegraphics[width=90mm]{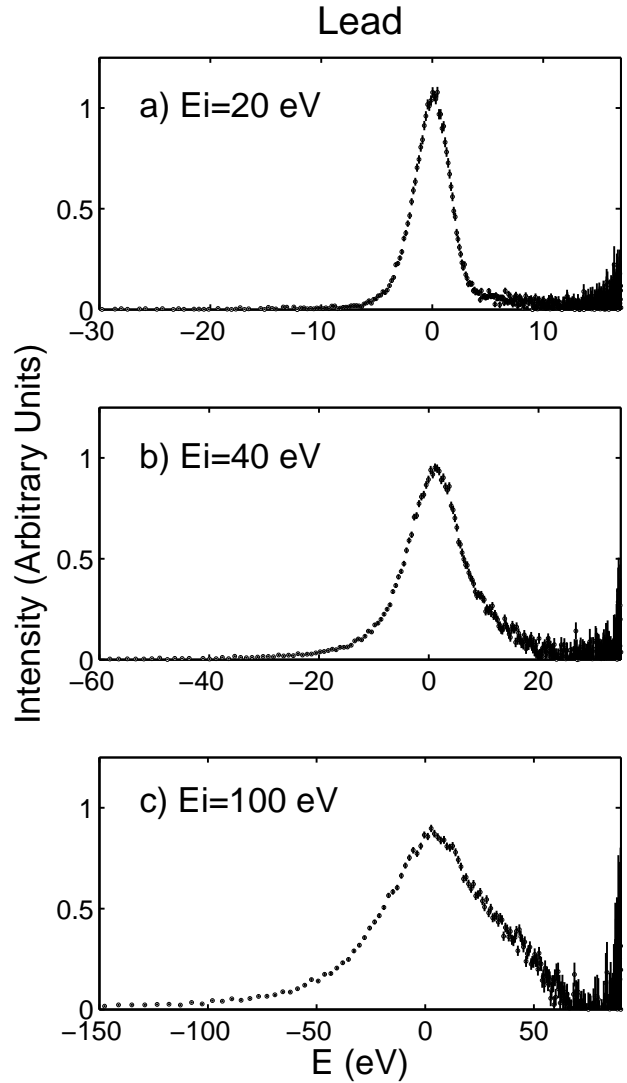}
\caption{The intensity as a function of energy transfer for a lead foil with E$_{i}$=20, 40, and 100 eV.  The data is the same as presented in Fig. \ref{lead_time}, though converted to an energy spectrum and corrected for the detector efficiency.  The data was obtained by integrating over $2\theta$=[30,70].} \label{lead_energy}
\end{figure}

\noindent Setting Eqn. \ref{cs1} equal to Eqn. \ref{cs2} gives the following expression.

\begin{eqnarray}
{S(Q,E)}=B E_{1}^{2} {{d^{2}\sigma}\over{d\Omega dt}}=B' \tilde{t}^{4} {{d^{2}\sigma}\over{d\Omega dt}},
\label{cs3}
\end{eqnarray} 

\noindent where $B$ and $B'$ are constants.  The expression assumes a direct geometry spectrometer with a fixed incident energy and a fixed phasing of the incident beam Fermi chopper.  The parameter $\tilde{t}$ is the time taken from when the neutron pulse hits the sample and when a signal is recorded on the detectors.  It should be noted that the parameter $t$ is the time from when the proton pulse strikes the target and these two different definitions of time will be maintained throughout this paper. Eqn. \ref{cs3} illustrates that $S(Q,E)$ is related to the measured intensity (corrected for detector efficiency) after a correction of $\tilde{t}^{4}$.  An example of the background subtracted data converted to an energy spectra is illustrated in Fig. \ref{lead_energy} for the data presented previously in Fig. \ref{lead_time} for a lead foil.  We note that the energy dependent detector efficiency was corrected for when the data was converted from time to energy.   

The width and the integrated intensity depends on the trajectory in the $E-Q$ plane of the measurement. So far in the discussion of the width we have assumed that we are using a constant wave vector trajectory and are varying only the energy. Since the data is collected by detectors at certain scattering angles it can be advantageous to use a constant scattering angle trajectory when problems arising from changing detectors across a scan do not arise. The disadvantage usually is that it is then necessary to use a Jacobian as first introduced by Waller and Froman.~\cite{Waller52:4}  For a conventional time-of-flight machine with hydrogen as the sample the Jacobian at the peak of the scattering is 

\begin{eqnarray}
J (2\theta) = 1 + {m \over M} \left(1 - {k_{0}\over k_{1}} \cos(2\theta) \right), 
\label{jac2}
\end{eqnarray} 

\noindent where $m$ is the mass of the neutron and $M$ is the mass of the sample atom.  For the case of hydrogen, $m$=$M$ and using this result in conjunction with the kinematical scattering expressions (Eqn. \ref{jac1}) we obtain that the Jacobian is identically unity. Consequently to a good approximation the intensity and width can be found directly by integrating the spectra at each scattering angle. The width is then the same as that obtained for a constant $\vec{Q}$ trajectory and the intensity has the same value as for a constant $\vec{Q}$ trajectory. 

\subsection{Impulse approximation and line shape}

To understand the observed cross sections and scattering profiles, we use the impulse approximation which is the high energy neutron scattering from a single atom averaged over all the different atoms in the sample.  The interference terms in the scattering amplitude are assumed to completely cancel. The scattering of neutrons with an incident energy of $E_{0}$ and wave vector $\vec{k}_{0}$ to a state with wave vector $\vec{k}_{1}$ and energy $E_{1}$ is given by 

\begin{eqnarray}
{I(E)}={A \exp \left( -{(E-C Q^{2})^{2} \over (WQ)^{2}} \right)}.
\label{lineshape}
\end{eqnarray} 

\noindent Where $E$ is the energy transfer $E_{0}-E_{1}$ and $Q$ is the wave-vector transfer $\vec{Q}=\vec{k}_{0}-\vec{k}_{1}$. The scattering function is normalised if $A = (\sqrt{\pi}WQ)^{-1}$ and the constant $C = \hbar^{2}/2M$ where we shall take $M$ as the mass of the hydrogen atom. The width of the scattering arises from the motion of the hydrogen atom immediately before the neutron collides with it and $W$ is proportional to the mean average velocity of the hydrogen atoms.  Eqn. \ref{lineshape} demonstrates that the width is proportional to momentum transfer $Q$ and hence the width of the recoil line is predicted to increase with increasing scattering angle and hence momentum transfer.

\begin{figure}[t]
\includegraphics[width=90mm]{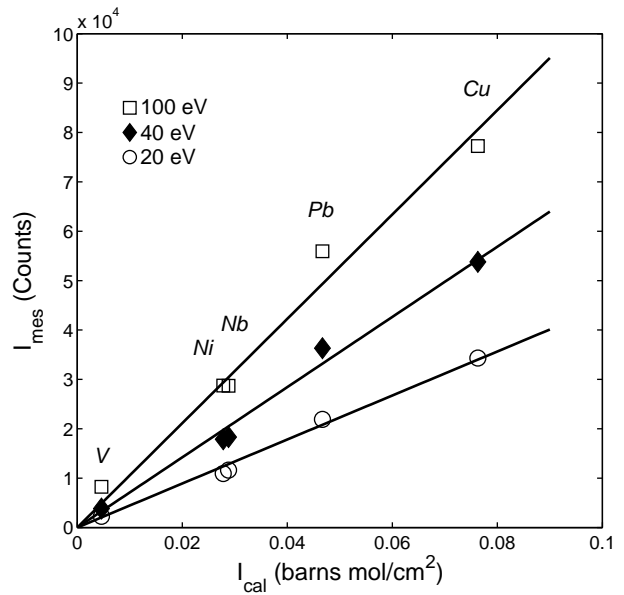}
\caption{The measured integrated intensity is plotted as a function of the calculated intensity per unit area for a series of calibration foils.  The thicknesses and cross sections are discussed in the text.  The intensities were obtained by integrating $2\theta$=[30,70].} \label{foils}
\end{figure}

It is useful to introduce dimensionless coordinates to describe the scattering by dividing the energies by the incident neutron energy $E_{0}$ and the wave vectors by the magnitude of the incident neutron wave vector $\vec{k}_{0}$. The expression for the scattering then becomes

\begin{eqnarray}
{I(\tilde{e})}={B \exp \left( -{(\tilde{e}-\tilde{q}^{2})^{2} \over (w\tilde{q})^{2}} \right)}.
\end{eqnarray} 

\noindent Where the reduced incident energy $\tilde{e} = E/E_{0}$ and $\tilde{q} = Q/k_{0}$.  The normalisation constant is $B = A/k_{0}$ and the reduced width is $w = W / (C k_{0})$. This expression shows that the peak energies for the scattering occur when $\tilde{e} = \tilde{q}^{2}$ and that this is independent of the incident neutron energy showing that a plot of the peak energies should be identical for all incident energies when plotted in terms of the reduced coordinates, $\tilde{e}$ and $\tilde{q}$. Rearranging the neutron scattering expressions for conservation of energy and momentum and using the expression for the peak of the scattering the wave vector transfer becomes $\tilde{q} = \sin(2\theta)$ where $2\theta$ is the scattering angle.  Again, this equation illustrates that the momentum transfer $\tilde{q}$ is always less than 90$^{\circ}$. 

The half-width of the scattering in these reduced coordinates is given in terms of the reduced energy as $\Gamma(\tilde{q})=\tilde{q} w \sqrt{\ln 2}$. This depends on the angle of scattering $\tilde{q}$ and is also proportional to the incident wave vector $k_{0}$. Hence as the incident energy increases $w$ decreases and the width in a reduced plot also decreases. Our conclusion is that when using these reduced coordinates the position of the peak should be independent of the incident energy but that the peak in the scattering should steadily get narrower as the incident energy increases.  Furthermore, if polyethylene can be approximated as a harmonic solid the line shape of the scattering is Gaussian and therefore we use this linsehape to investigate the recoil scattering in detail.  

\section{Intensity calibration and absolute units}

To convert the measured intensities to an absolute cross section with units of barns, a series of standard materials with known cross-sections were compared.  We have chosen to use foils of copper, lead, nickel, niobium, and vanadium.  The thicknesses of the metal foils are listed in Table \ref{table_foils} and were chosen to scatter less than 10 \% of the incident beam so that corrections due to multiple scattering would not need to be considered.  The use of heavy atom foils with masses much larger than the hydrogen atom means that the recoil energy was small in comparison to both the recoil energy of the hydrogen atom and the experimental resolution. 

The cross sections, $\sigma$ were taken from the tables compiled by Sears (Ref. \onlinecite{Sears92:3}) for neutron energies of 25.3 meV.  The metal foils chosen have cross sections which vary little over the energy range of interest and do not display strong resonances below $\sim$ 100 eV.  

\begin{table}[ht]
\caption{Foil Standards}
\centering
\begin{tabular} {c c c c c}
\hline\hline
X & d (mm) & $\rho$ (g/cm$^{3}$) & $\sigma$ (barns) & $m$ (g/mol) \\
\hline\hline
V & 0.085 & 6.11 & 5.1 & 50.94 \\
Ni & 0.11 & 8.908 & 18.5 & 58.69 \\
Nb & 0.51 & 8.57 & 6.3 & 92.91 \\
Pb & 0.86 & 11.34 & 11.1 & 207.2 \\
Cu & 0.73 & 8.94 & 8.0 & 63.55 \\
\hline
\label{table_foils}
\end{tabular}
\end{table}

\begin{figure}[t]
\includegraphics[width=85mm]{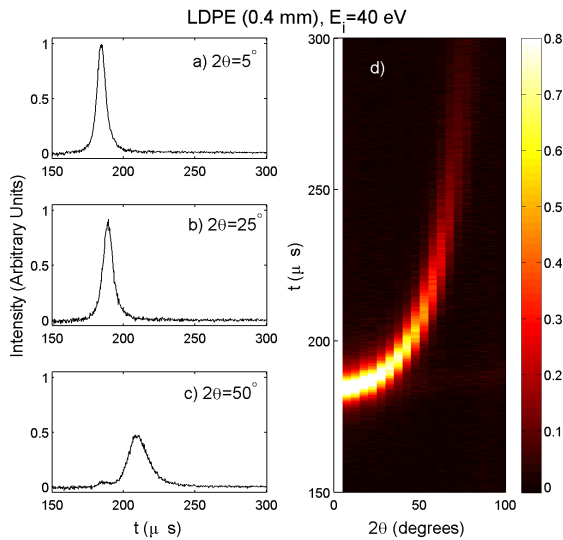}
\caption{The scattering from a 0.4 mm thick film of LDPE as a function of time of flight.  Panels $a)-c)$ represent constant 2$\theta$ cuts at 5$^{\circ}$, 25$^{\circ}$, and 50$^{\circ}$ respectively.  Panel $d)$ is a false color plot of the intensity as a function of $2\theta$ and $E$.} \label{time_poly}
\end{figure}

From the values in Table \ref{table_foils} we calculated an expected cross section per unit area for each foil (in units of $barns \times mol/cm^{2}$) equal to $I_{cal}=d \rho \sigma /m$ where $d$ is the thickness, $\rho$ is the density, and $m$ is the molar mass.  The measured intensity (integrated over $2\theta$=[30$^{\circ}$,70$^{\circ}$]) was then plotted against the calculated cross section to obtain a calibration constant $\alpha_{foils}$ from the slope.  The calibration curves for E$_{i}$=20, 40, and 100 eV are illustrated in Fig. \ref{foils}.  The measured intensity scales linearly with the calculated intensity and therefore confirms that multiple scattering effects do not need to be considered.  The systematic increase in slope with incident energy represents the fact that the flux of neutrons increases with incident energy.

\begin{figure}[t]
\includegraphics[width=95mm]{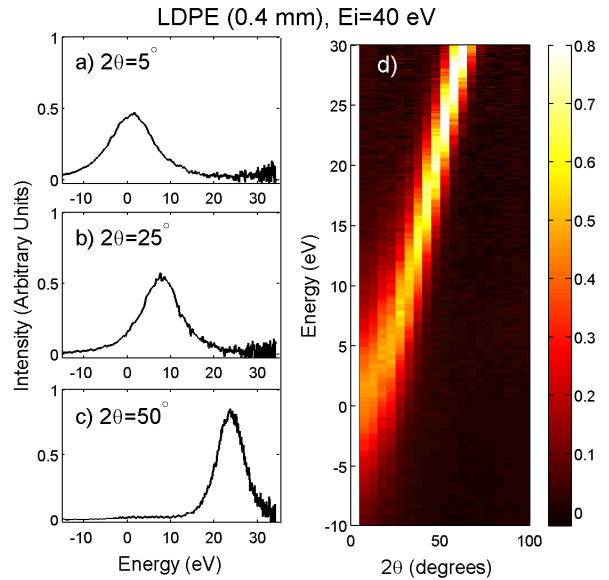}
\caption{The scattering from a 0.4 mm thick film of LDPE as a function of energy transfer.  Panels $a)-c)$ represent constant 2$\theta$ cuts at 5$^{\circ}$, 25$^{\circ}$, and 50$^{\circ}$ respectively.  Panel $d)$ is a false color plot of the intensity as a function of $2\theta$ and $E$.} \label{LDPE_energy}
\end{figure}

\begin{figure}[t]
\includegraphics[width=90mm]{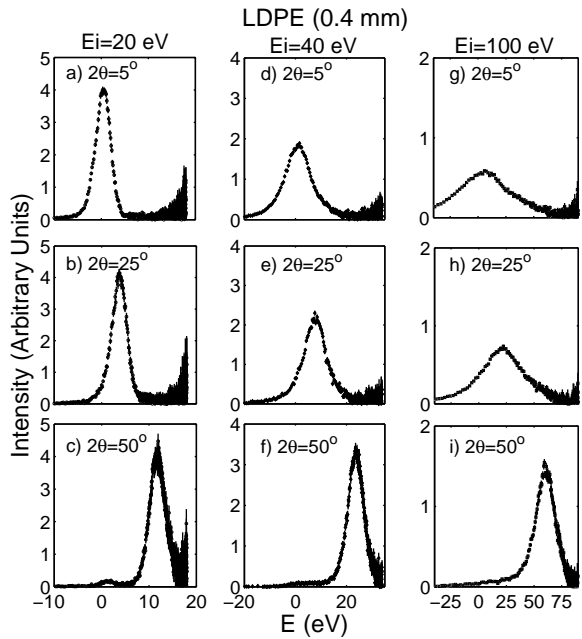}
\caption{The measured intensity plotted as a function of energy transfer with a sample of LDPE (0.4 mm).  Panels $a)-c)$ are for E$_{i}$=20 eV,$d)-f)$ are for E$_{i}$=40 eV, $g)-i)$ are for E$_{i}$=100 eV.  The data is plotted for representative scattering angles of 5$^{\circ}$, 25$^{\circ}$, and 50$^{\circ}$.} \label{summary_poly}
\end{figure}

The curves provide an accurate means of putting the measured integrated intensity on an absolute scale.  The increased scatter in the data points for higher incident energies is possibly due to uncertainty in the energy dependence of the cross section.~\cite{Barn_Book}  While the energy dependence of the cross sections have been tabulated, we have found substantial variations in the tabulated values (up to $\sim$ 10\%) and therefore have chosen to take the values at 25.3 meV which have been measured and carefully tabulated.  These values should therefore provide a good approximation of the bound cross section. Based on the slopes ($\alpha_{foils}$) obtained from Fig. \ref{foils}, the integrated intensity can be converted to an absolute cross section in units of barns. We consider that a large component of the error is due to uncertainty in the energy dependence of the cross-sections in the range 1-100 eV.  Improved measurements are needed to obtain more accurate results.

\section{Hydrogen scattering}

In the previous sections we outlined the background subtraction, conversion of the time spectra to energy, and the calibration of the spectrometer using a series of metal foils.  In this section we now discuss the results for polyethylene in both units of time-of-flight and energy transfer.  We then discuss the momentum dependence of the hydrogen recoil line line shape including the recoil peak position, width and integrated intensity.  

\subsection{Hydrogen recoil in polyethylene}

\begin{figure}[t]
\includegraphics[width=90mm]{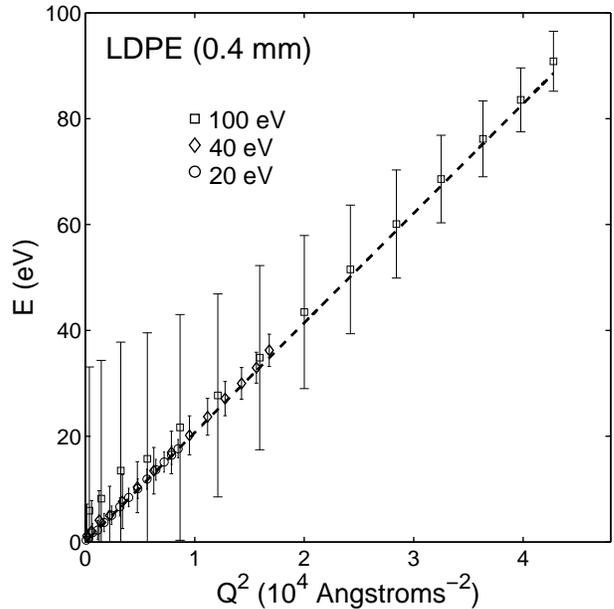}
\caption{The position of the hydrogen recoil line is plotted as a function of $Q^{2}$ for E$_{i}$=20, 40, and 100 eV.  The dashed line is $E=\hbar^{2}Q^{2}/2M_{p}$.  The errorbars represent the full width at half maximum intensity of the hydrogen recoil peak.  The fitted error bars are approximately the size of the data points.} \label{position}
\end{figure}

Representative time-of-flight spectra for the LDPE sample ($d$=0.4 mm) are illustrated in Fig. \ref{time_poly}.   The data in panels $a-c)$ were obtained with an incident energy of E$_{i}$=40 eV at three different scattering angles (5$^{\circ}$, 25$^{\circ}$, and 50$^{\circ}$ respectively) sampled at an equal time intervals of 0.25 $\mu s$.  Panel $d)$ illustrates a false contour plot of the intensity as a function of both time-of-flight and scattering angle.  The energy converted data is displayed in Fig. \ref{LDPE_energy} and is based on the same data set as in Fig. \ref{time_poly} using the procedure outlined in the previous sections.  Large energy transfers corresponds to longer time and since there are more time bins per unit energy transfer, the density of points increases with energy transfer.  

At the lowest scattering angle ($2\theta$=5$^{\circ}$ in panel $a)$ of Fig. \ref{time_poly} and Fig. \ref{LDPE_energy}), the momentum transfer is sufficiently small that the recoil lines of hydrogen and carbon overlap and appear at the elastic ($E$=0 in Fig. \ref{LDPE_energy}) position (appearing at a time of $\sim$ 180 $\mu s$ in Fig. \ref{time_poly}).  Since the incident energy is fixed on MARI, increasing energy transfer correponds to larger times on the plots illustrated in contrast to indirect geometry spectrometers where energy transfer grows with decreasing time-of-flight.  

As the scattering angle increases along with the momentum transfer, we observe a gradual shift of the recoil line ($2\theta$=25$^{\circ}$) and then finally a separation of the carbon and hydrogen recoil lines clearly observed at $2\theta$=50$^{\circ}$ with the carbon near $E$=0 and the hydrogen recoil peak appearing at finite energy transfer. Because the energy dependence of the recoil line for an atom of mass $m$ is $E=\hbar^{2}Q^{2}/2m$, the carbon atom will recoil with a smaller energy than the hydrogen atom for a given momentum transfer.  The recoil peak of carbon will therefore appear at smaller energy transfers, or at shorter times given the geometry of our time-of-flight instrument.

A series of different representative constant $2\theta$ cuts for the three different incident energies studied is illustrated in Fig. \ref{summary_poly} for scattering angles of $2\theta=5^{\circ}$, $25^{\circ}$, and $50^{\circ}$.  All three different incident energies display the same qualitative features with a strong hydrogen recoil peak which shifts in energy as a function of $2\theta$ and a very weak carbon recoil peak located near the elastic position at all scattering angles.   

\subsection{Variation of Scattering with Q}

\begin{figure}[t]
\includegraphics[width=90mm]{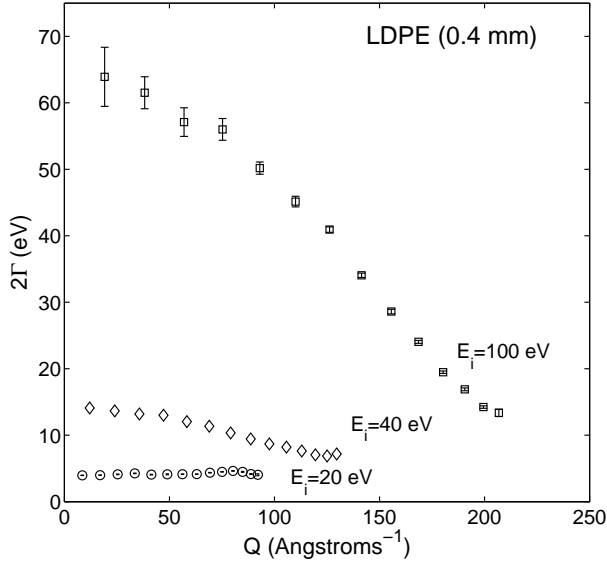}
\caption{The full-width as a function of momentum transfer is plotted for E$_{i}$=20, 40, and 100 eV.  The data is from a 0.4 mm thick film of LDPE.} \label{width}
\end{figure}

The momentum dependence of the hydrogen recoil line was obtained by fitting the constant $2\theta$ scans to the sum of two Gaussians to represent the recoil lines from hydrogen and carbon with a total of 6 parameters (2 amplitudes, linewidths, and positions).  To restrain the fit and obtain an accurate measure of the hydrogen recoil position at small scattering angles, where both the carbon and hydrogen recoil lines overlap, we fixed the amplitude and linewidth of the carbon line to be that measured at scattering angles greater than 100$^{\circ}$.  Because the neutron and proton masses are approximately equal, there is no  hydrogen recoil above a scattering angle of 90$^{\circ}$.  With these constraints, we were able to obtain a good description of the line shape and scattering angle dependence for the three different incident energies of the neutrons.

The peak position of the hydrogen recoil line is displayed in Fig. \ref{position} with the errorbars representing the measured width for the different incident neutron energies used. According to the impulse approximation the peak position should scale with momentum transfer $Q$ as $E=\hbar^{2}Q^{2}/2M_{p}$ and is represented by the dashed line in Fig. \ref{position}.  The agreement between experiment and theory is excellent over the entire range of $Q$ and for the different incident energies.  We observe no deviation which would be expected if there was a breakdown of the Born Oppenhemier approximation predicted by several theories.~\cite{Gido05:71}

The full-width of the scattering ($2\Gamma$) is shown in Fig. \ref{width}.  The results in contrast to expectations based on the impulse approximation which predict an increase in the linewidth with increasing momentum transfer shows a decrease.  This figure makes clear that the width is largely dominated by the experimental resolution which, particularly for incident energies of 100 eV, becomes much smaller at larger $Q$. As discussed in Appendix B, this is because it is dominated by the width of the incident neutron pulse from the target and moderator.  With 20 eV incident neutrons this is less of a problem and there is a slight increase in the full-width with increasing $Q$.  At low angles the full width of the resolution is approximately 15 \%, 35 \%, and 65 \% for incident energies of 20 eV, 40 eV, and 100 eV respectively and at large angles is 25 \%, 17 \%, and 14 \% at high angles.

\begin{figure}[t]
\includegraphics[width=80mm]{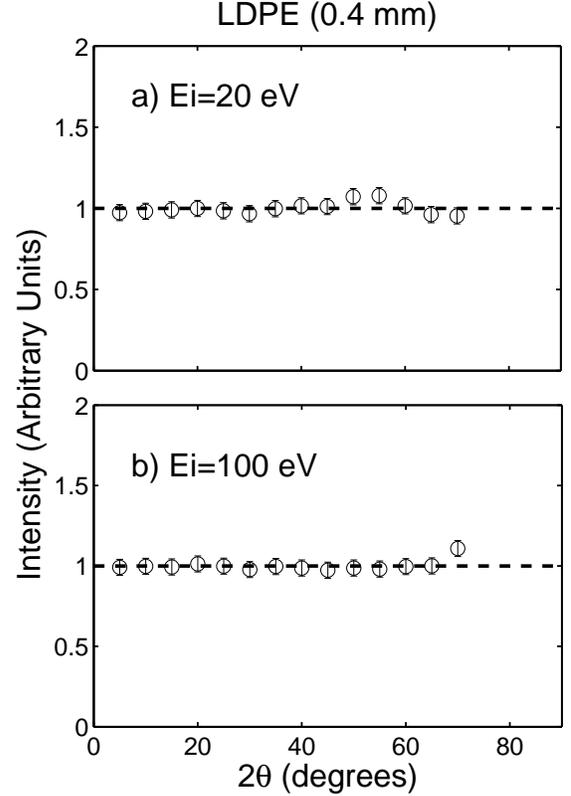}
\caption{The integrated intensity of the hydrogen recoil line as a function of $2\theta$ for $a)$ E$_{i}$=20eV and $b)$ E$_{i}$=100 eV.} \label{intensity_separate}
\end{figure}

\begin{figure}[t]
\includegraphics[width=90mm]{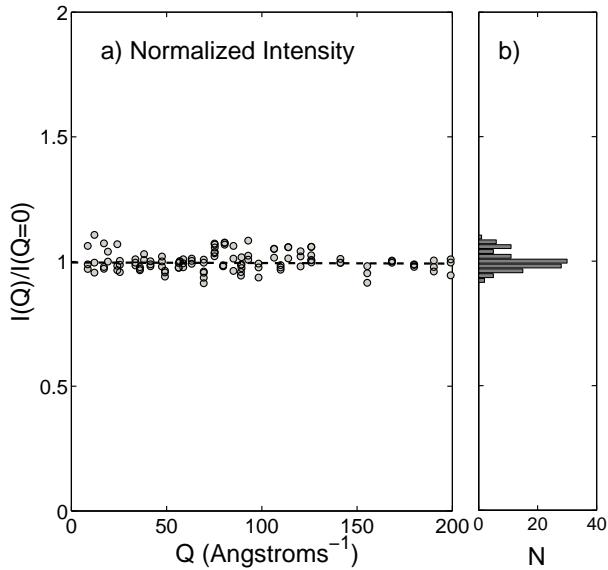}
\caption{$a)$ The intensity normalised as  $Q \to 0$ is plotted as a function of momentum transfer for E$_{i}$=20, 40, and 100 eV.  The mean of the data is 1.00 with a standard deviation of 0.04.  $b)$ A histogram of the of the observed intensities is plotted illustrating the fluctuations of the data.} \label{norm_intensity}
\end{figure}

The integrated intensity as a function of scattering angle is plotted in Fig. \ref{intensity_separate} for the LDPE ($d$=0.4 mm) for E$_{i}$=20 (panel $a)$) and 100 eV (panel $b)$).  The data sets represent the two extreme ranges of momentum transfer investigated and illustrate no measurable momentum dependence to the integrated intensity of the hydrogen recoil line.  A summary of all of the data taken at each incident energy as a function of momentum transfer ($Q$) is plotted in Fig. \ref{norm_intensity}.  The data in both Figs. \ref{intensity_separate} and \ref{norm_intensity} have been obtained by dividing by the intensity for each incident neutron energy and sample extrapolated to its value at $Q$=0.   The density of data points at lower momentum transfers in Fig. \ref{norm_intensity} is larger owing to the significant overlap of data points from each incident energy at low scattering angles.   For the largest momentum transfers, only the E$_{i}$=100 eV data set contributes.  Based on the data in panel $a)$ we find no momentum dependence with the dashed line equal to 

\begin{eqnarray}
{I(Q)\over I(Q=0)}=1.00 \pm 0.04.
\end{eqnarray} 

\noindent The error is derived from the standard deviation of the data points in Fig. \ref{norm_intensity} $a)$.  Panel $b)$ is a histrogram of the points in panel $a)$ and visually illustrates the statistical spread of the data.  Based on the combined data in Fig. \ref{norm_intensity}, we conclude there is no momentum dependence to hydrogen cross section.  This is in contrast with previous results taken on indirect geometry machines (as illustrated in Fig. \ref{water}) and most notably those reported in Ref. \onlinecite{Cowley06:18} which spanned a momentum transfer of up to $\sim$ 100 \AA$^{-1}$.

\subsection{Absolute cross section}

Having shown that the intensity of the hydrogen recoil scattering is constant with momentum transfers ranging up $\sim$ 250 \AA$^{-1}$, we now discuss the absolute value for the hydrogen cross section.  We have obtained 9 independent measurements (three different densities at three different incident energies) of the recoil cross section in polyethylene and each measurement can be put on an absolute scale through the calibration curves obtained from the measurement of metal foils described above.

\begin{figure}[t]
\includegraphics[width=85mm]{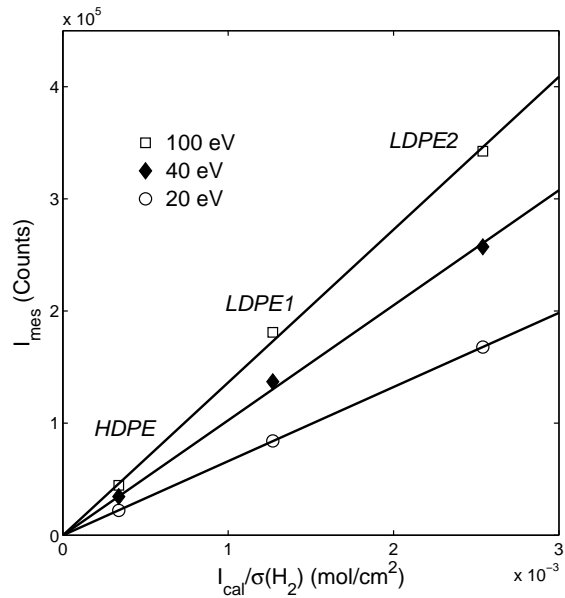}
\caption{The measured intensity as a function of calculated intensity normalised by the number of hydorgen atoms.} \label{poly_summary}
\end{figure}

Figure \ref{poly_summary} summarizes the results for the three different polyethyelene samples measured at the three different incident energies.  The calculated intensity is normalised by the hydrogen cross section in each formula unity of polyethylene (CH$_{2}$) and therefore represents a measure of the number of hydrogen atoms per unit surface area.  The measured intensities for each density and thickness vary linearly with the density and therefore illustrate that multiple scattering processes are negligible.

\begin{figure}[t]
\includegraphics[width=90mm]{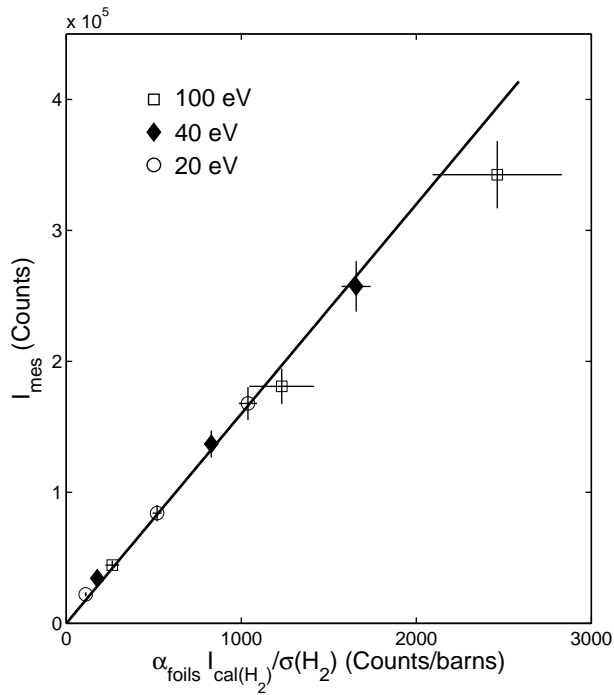}
\caption{A plot of the measured intensity as a function of the calculated intensity as described in the text.  All three different incident energies and the three thickness of HDPE and LDPE are plotted and denoted by different symbols.  The slope of the curve is the absolute cross section of $\sigma(H_{2})$.} \label{abs_H}
\end{figure}

The momentum dependence was found to be independent of the scattering angle for each density of polyethyelene at each incident energy. The average intensity from each data set gives a different measurement of the recoil cross section and there are 3 polyethylene samples and 3 different incident neutron energies giving 9 different measurements of the average intensity.  The intensity of each measurement ($I_{mes(H_{2})}$ in units of 'Counts') was plotted against the value $\alpha_{foils} \times I_{cal(H_{2})}/\sigma(H_{2})$ (in units of Counts/barns) represented in Fig. \ref{poly_summary} and the result is showin in Fig. \ref{abs_H}.  The straight line has a slope which is the hydrogen cross section in one formula unit of polyethyelene denoted $\sigma(H_{2}$). 

The results are displayed in Fig. \ref{abs_H} and illustrate that all of the data do fall on a straight line illustrating the consistency of each data set obtained with different incident energies and differing energy resolutions. The slope of the line divided by 2 is the bound cross section of a single hydrogen atom,

\begin{eqnarray}
\sigma(H)=80 \pm 4\ barns,
\end{eqnarray} 

\noindent in perfect agreement with value of 82.02 $\pm$ 0.06 quoted by Sears (Ref. \onlinecite{Sears92:3}) for the bound cross section at an energy of 25.3 meV.  The error bars in Fig. \ref{abs_H} and the resulting error bar on the final absolute cross section is the result of uncertainties in the foil thickness, foil cross section at high energies, and also the thickness and hence density of polyethyelene films.  Nevertheless, our data is in good agreement with the expected cross section based on the impulse approximation.

In comparison, previous results based on indirect spectrometers have given values ranging from $\sigma(H)$=48-64 barns at large momentum transfers of about 100 \AA$^{-1}$.  The experiment discussed here clearly rules this value out over a much broader range in momentum transfer.

\section{Conclusions and discussion}

The experiment described above has shown conclusively three key results.  Firstly, a direct geometry time-of-flight machine can be used with incident energies up to at least 100 eV and the background quantitatively subtracted.  Secondly, the hydrogen recoil cross section is constant over the entire momentum transfer range of the experiment ($Q\sim250 \AA^{-1}$).  Thirdly, the absolute value of the cross section is in perfect agreement with previous measurements of the bound cross section as expected from conventional scattering theory.  All of these results are in contrast to the results discussed in the introduction on a variety of hydrogen containing systems.  

While we have only conducted experiments on polyethylene, the results are much more general.  The impulse approximation assumes that the scattering sites are independent and therefore the results are indeed general to all hydrogen containing systems.  Hence, our results stand in contrast to the variety of hydrogen containing systems where a deficit in intensity was observed.

It is also of interest to discuss and compare the energy widths of our data and those obtained using Vesuvio when the cross section was anomalous.  In Fig. \ref{width} we have presented all of our data in terms of energy transfer, while previous results were presented in terms of time.  The expression relating energy resolution to the timing resolution can be obtained by considering a fixed scattering angle $2\theta$ and a spectrometer configuration in which $E=C/t^{2}$. Taking derivatives of this expression we obtain $\Delta E/E \sim 2 \Delta t/t$.  Using the results of Fig. \ref{width} the energy resolution on MARI at large scattering angles is $\Delta E/E \sim 20 \%$. The experimental data shown in Fig. \ref{water} gives an energy resolution of about $2 \Delta t/t \sim 50 \%$.  A comparable data set for polyethylene on Vesuvio (Ref. \onlinecite{Cowley06:18}) has been analyzed and the energy widths obtained are all substantially larger than the widths shown in Fig. \ref{width} except at the lowest energies where the resolution in the MARI experiment could easily be improved by using a lower incident energy.  When, however, the experimental widths are corrected for the Waller-Froman factor (Ref. \onlinecite{Waller52:4}) this decreases the width in the indirect geometry experiment by a large factor especially at the highest scattering angles.  The minimum energy widths of both experiments are then very similar. Therefore unlike previous results, our data cannot be reconciled with the indirect spectrometer data based upon the experiments having a different energy resolution.


For Q=100 \AA$^{-1}$, experiments on polyethylene on Vesuvio have obtained an energy width of $\sim$ 25 eV where we measure $\sim$ 7 eV.  If we divide by the Waller-Froman Jacobain discussed earlier, the VESUVIO hydrogen width becomes 5 eV, comparable with the MARI results presented here.  Therefore, depending on how the data is analysed, the hydrogen recoil widths have been measured here to be either better or at least comparable to studies on Vesuvio.  We note that corrections for non Gaussian line shapes were found not to alter the conclusions obtained fron indirect geometry spectrometers.~\cite{Mayers04:16}  Unlike previous results, we conclude that our data cannot be reconciled with the indirect spectrometer data based on a differing energy resolution.  

Through the use of several different incident energies with very different resolutions, we have shown that we obtain consistent answers both for the absolute cross section for hydrogen and for the constant momentum dependence.  Therefore, the results do not depend on energy resolution as claimed in Refs. \onlinecite{Mayers07:98,Dreis00:84}.  The experimental resolution would be important if there was some breakdown in the Born Oppenheimer approximation when the line shape would be distorted by interaction with the electronic energy levels.  We suggest that these are unlikely to be observed and to cause strong deviations except where there is a strong coupling to a low energy mode possibly of order 100-500 meV or less.  We have not observed any deviations in the line shape from polyethylene even though several electronic calculations have found energy levels with separation $\sim$ 10 eV(Refs. \onlinecite{Miao96:54,Falk73:6}) which are in good agreement with photoelectron measurements (Ref. \onlinecite{Wood72:56}).  

We therefore conclude that the discrepancy in results from this experiment and the previous data based on indirect geometry machines is due to experimental error.  While this point needs to investigated through a series of careful experimental checks, Ref. \onlinecite{Cowley06:18} has suggested that understanding the flux distribution of the incident beam maybe one possibility for the problem. Another possibility could be the background introduced through the presence of a large amount of high-energy radiation from the initial interaction of the proton pulse with the tungsten target.  In conventional experiments on direct geometry instruments such as MARI, MAPS, and MERLIN, these high-energy particles are removed through the use of a nimonic chopper.  Such a chopper does not exist on indirect geometry instruments and was not used in this experiment to access the high energy spectrum. While the background at short times is surprisingly large, we have found an excellent way to subtract this background that is comparable to the signal at incident energies of at least up to 100 eV.  Presumably this background is also present on the indirect spectrometers but it is impossible to subtract this background on these machines.  Such a background may not have been as severe in the case of the electron linac experiments discussed in Ref. \onlinecite{Blostein09:102,Devereaux07:79}.  This point will need to be investigated in future work.

While our results show conclusively the lack of a deficit and anomalous cross section, the experiments do show that direct geometry instruments can be used for large energy transfers.  Previously, the use of direct geometry have been restricted to energies below $\sim$ 500 meV (see for example Refs. \onlinecite{Zaliz04:93,Stock07:75,Coldea01:86}) to study magnetic and electronic excitations.  There has been a considerable amount of interest in excitations at energy transfer greater than $\sim$ 1 eV.  These have mostly been pursued with inelastic x-rays and light scattering where the condition of large energy transfers and small momentum transfers can be simultaneously achieved.~\cite{Bruch07:79,Devereaux07:79,Kotani01:73}  It will be a topic of future study to investigate whether neutrons can be used to study high energy magnetic and electronic excitations at energy transfers greater than $\sim$ 1 eV. 

In conclusion, we have used a direct geometry chopper spectrometer to measure the recoil cross section of hydrogen in polyethylene.  We find the cross section to be independent of momentum transfer, in perfect agreement with values measured by Sears, and the results to be independent of energy resolution.  The data are in good agreement with theory based on the impulse approximation.  All of these results are in contrast to previous experiments on indirect geometry spectrometers and with theories based on quantum entanglement.  We suggest that the discrepancy is the result of experimental issues using indirect geometry spectrometers.  Our results do show the use and feasibility of direct geometry spectrometers at very large energies and we suggest future possible uses for these spectrometers.

\begin{acknowledgements}

We would like to thank K. Allen for expert help with the choppers; A. Orszulik for machining and design of the sample holders; E. Schooneveld and N. Rhodes for help with the detectors and monitors; and N. Gidopoulos, J. Mayers and J. Tomkinson for scientific discussions. R. A. C. is grateful for support from the Lindemann Trust and STFC.

\end{acknowledgements}

\section{Appendix A: Energy calibration}

For energies where MARI is typically used (below $\sim$ 500 meV), the energy is calibrated from the time the neutron pulse reaches $Mon$ 2 located just after the Fermi chopper (Fig. \ref{spectrometers}).   Because the monitors are sensitive to $\gamma$ radiation (which are normally removed by the nimonic chopper) the short time spectrum of the monitors could not be used and therefore this method could not be relied upon for calibrating high incident energies beyond $\sim$ 40 eV.   We therefore relied upon $Mon$ 3 which is located further down the beam path implying it views less of the initial $\gamma$ flash from the protons striking the target and also the time of arrival of the neutrons will occur at considerably later times than the initial high energy background.  Therefore, for our energy calibration and the low-angle detectors to calibrate the incident energy.

\begin{figure}[t]
\includegraphics[width=80mm]{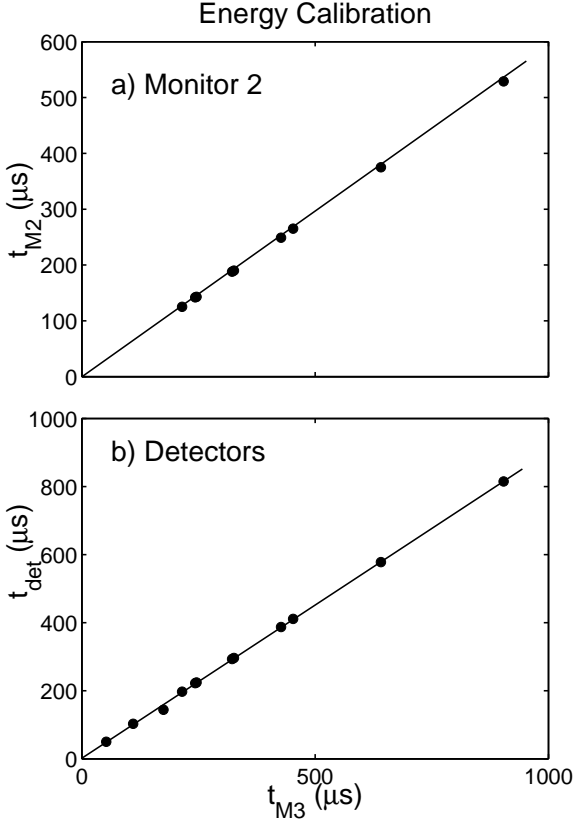}
\caption{$a)$ The time of arrival of the neutron pulse at $Mon$2 as a function of time of arrival at $Mon$3.  $b)$ The time of arrival of the neutron at the low-angle detectors spanning $2\theta$=[5$^{\circ}$,7$^{\circ}$] as a function of time of arrival at $Mon$3.} \label{det}
\end{figure}

Because the detectors and the monitors are based on different electronics, it is important to calibrate them against one another to ensure there are no systematic time differences.    Fig. \ref{det} $a)$ illustrates a plot of the time of arrival of the neutron pulse at $Mon$2 as a function of time of arrival at $Mon$3.  The data only includes spectrum from incident energies less than $\sim$ 40 eV as the time-of-flight peak position in $Mon$2 becomes difficult to reliably determine due to the background caused by the $\gamma$ radiation normally removed by the nimonic chopper.  The slope is the measured path length difference between the two monitors ($L_{Mon3}/L_{Mon2}$) and describes the data very well implying that $Mon$3 can be used to calibrate the energy of the incoming neutron pulse for higher incident energies.  This curve illustrates that the calibration on $Mon$3 is consistent and can be reliably compared with $Mon$2 at high incident energies or short times of flight. 

To test the electronics of the detectors and to ensure the time-of-flight of arrival is consistent for both He$^{3}$ detectors and glass bead monitors, Fig. \ref{det} $b)$ displays the time of arrival at the low-angle detectors with a vanadium sample integrating over detectors ranging spanning $2\theta$=[5$^{\circ}$,7$^{\circ}$].  The detectors are located at sufficiently small momentum transfer that we expect that (for the case of vanadium) that no measurable recoil energy will be observable and the scattering will be dominated by the elastic incoherent cross section of vanadium.  The fit in Fig. \ref{det} $b)$ is to a line with a slope equal to the ratio of the path lengths ($L_{det}/L_{Mon2}$) and with an intercept of 3.5 $\mu s$.  The intercept implies that the detector electronics count later than the monitors and the time difference must be corrected in calibrating the energy of the neutrons, particular at high energies where the experiment occurs at very short times ($\sim$ 100-200 $\mu s$).  Based on the data presented in Fig. \ref{det}, we are able to calibrate the incident neutron energy at high-energies (or very short times) and also the energy transfer of the detected neutrons.  We hope that these studies will be useful in any future high-energy experiments on direct geometry spectrometers.

\section{Appendix B: Experimental Resolution}

\begin{figure}[t]
\includegraphics[width=80mm]{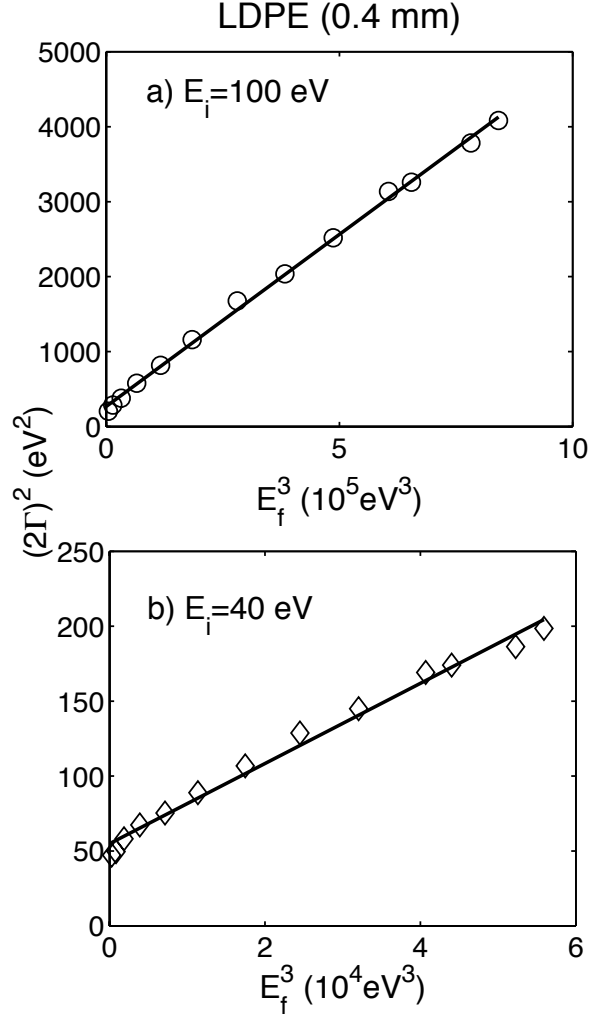}
\caption{$a)$ The full-width in energy squared ($2\Gamma^{2}$) of the hydrogen recoil line is plotted as a function of the final energy cubed ($E_{f}^{3}$) for E$_{i}$=100 eV (panel $a$) and E$_{i}$=40 eV (panel $b$).  The fit and parameters are described in the text.} \label{width_appendix}
\end{figure}

The line-widths of the hydrogen recoil lines in energy are shown in Fig. \ref{width} for the different incident neutron energies and how they vary over the range of angles which we expect to be valid in our experiment.  An unexpected feature of these results is the marked decrease in the line-width with scattering angle for large incident energies (a factor of $\sim$ 3.5 for an E$_{i}$=100 eV). This result is surprising because, as discussed above, the width of the hydrogen scattering is expected to increase with increasing angle in line with the impulse approximation whereas the experimental results show a steady decrease for large incident energies and large momentum transfers.  Furthermore, it is commonly expected that the neutrons are produced in a very short time, about 0.4 $\mu sec$ and the chopper then takes a slice of incident energies by being open for a considerably longer time. The uncertainty in the incident energy (determined by the Fermi chopper opening) is then expected to dominate the energy resolution for all energies. This is clearly not the case in this experiment especially for incident neutron energies of 100 eV and we will now consider this contradiction in more detail.

In principle there are two substantial contributions to the energy width of a direct geometry time-of-flight spectrometer. The neutrons are emitted by the moderator in a short pulse first which is about 0.5 $\mu sec$ long and this will be treated as instantaneous. If the chopper is open for a time $\tau_{\circ}$ then it will allow through all those neutrons which have the appropriate speeds to be able to pass through the chopper.  These neutrons will have a range of velocities and that is given in terms of the energy by,

\begin{eqnarray}
{{\Delta E_{\circ}} \over {E_{\circ}}}= {{2 \tau_{\circ}} \over L_{\circ}} \sqrt{2E_{\circ} \over m},
\end{eqnarray}

\noindent where $E_{\circ}$ is the incident neutron energy, and $L_{\circ}$ is the distance between the moderator and the chopper. This becomes 

\begin{eqnarray}
{{\Delta E_{\circ}} \over {E_{\circ}}}=  0.0027 \tau_{\circ} \sqrt{E_{\circ}}
\end{eqnarray}

\noindent when the incident energy is expressed in eV, the time $\tau_{\circ}$ in $\mu sec$ and the distance in meters for the Mari direct geometry spectrometer.  This energy width is then almost independent of the scattering angle apart from some smaller factors such as the thickness of the sample.

The other effect that arises is that due to the neutrons having different velocities because the slower neutrons will arrive at the sample position after the faster neutrons. This implies that in addition to the width coming from the different energies there is an additional width due to the spreading of the pulse with time. We shall assume that the spreading in time is given by $\tilde{\tau}$.  A simple argument to estimate this contribution to the width is to estimate it as the opening time of the chopper multiplied by the distance from the moderator to the sample divided by the distance of the chopper from the moderator.  This gives $\tilde{\tau}=1.17 \tau_{\circ}$.  The neutrons have a definite time delay at the target which they continue to have throughout the instrument. However, the neutrons are recorded in time channels and to compare with the predictions in Fig. \ref{width} the spectra must be converted into energy spectra. We therefore relate the lengths of the neutron pulse in time to the final energy E$_{1}$ by the following transformation: 

\begin{eqnarray}
{{\Delta E_{1}} \over {E_{1}}}=  {2 \tilde{\tau} \over L_{1}} \sqrt{2 E_{1} \over m}.
\end{eqnarray}

\noindent When the energies are in eV, the length in metres and the time in $\mu sec$ this becomes

\begin{eqnarray}
{{\Delta E_{1}} \over {E_{1}}}=  0.0067 \tilde{\tau} \sqrt{E_{1}}.
\end{eqnarray}

\noindent Assuming that these two contributions to the width of the scattering are independent we obtain the total width ($\Delta E$) at the detector as

\begin{eqnarray}
\Delta E=\sqrt{\Delta E_{\circ}^{2}+\Delta E_{1}^{2}}.
\end{eqnarray}

\noindent This analysis can be tested against the experiment by plotting the full-width ($2\Gamma$) against the final neutron energy E$_{1}$.  In particular, a plot of $(2\Gamma)^{2}$ as a function of $E_{1}^{3}$ gives the following expression for a straight line

\begin{eqnarray}
(2\Gamma)^{2}=\alpha + \beta E_{1}^{3}.
\end{eqnarray}

\noindent with $\alpha \equiv (0.0027 \tau_{\circ})^{2} E_{\circ}^{3}$ and $\beta \equiv (0.0067 \tilde{\tau})^{2}$.  The intercept and slope therefore provide a measure of $\tau_{\circ}$ and $\tilde{\tau}$ respectively.  The data and fit plotted in this manner are presented in Fig. \ref{width_appendix}.

This expression gives, at least approximately, an excellent description of the experimental data with 100 eV incident neutrons if $\tilde{\tau}$= 10 $\mu sec$ and $\tau_{\circ}$=6 $\mu sec$.  It also correctly describes the data taken with E$_{i}$=40 eV with $\tilde{\tau}$= 8 $\mu sec$ and $\tau_{\circ}$=11 $\mu sec$.  The times fitted from the 100 eV data have a ratio of 1.67 which is somewhat larger than predicted by the simple model described above. The data at 40 eV have a ratio which is smaller and the discrepancy indicates there are other contributing factors to the broadening of the time pulse.  Possibilities include sample size, increasing transmission of the boron blades of the chopper with increasing incident energy, and error in the phase of the Fermi chopper.  Despite these factors, the angular dependent widths are well described by our simple model governed by two time scales. 

We consider that this qualitative explanation of the resolution function is correct and that a full calculation including the size and shape of the scattering elements and the thickness of the detectors as well as dropping the independence of the time and energy aspects of the resolution would produce better answers. However, we do not wish to follow this here as we are confident that we have identified the main components of the resolution.


\end{document}